Title:

# Exact calculation of single-electron states in Si-nanocrystal embedded in $SiO_2$

Authors:

**Nguyen Hong Quang**[a,b], **Tran The Trung**[a,c], **Johann Sée**[b] **and Philippe Dollfus**[b,*]

[a] Institute of Physics, NCST, 46 Nguyen Van Ngoc, Hanoi, Vietnam

[b] Institut d'Electronique Fondamentale, Bâtiment 220,
UMR 8622 – CNRS – Université Paris-Sud XI, 91405 Orsay, France

[c] Service d'Aéronomie, Réduit de Verrières,
B.P.3, Route des Gâtines, 91371 Verrières le Buisson, France

## Abstract

We present an exact calculation of the single-electron energies and wave-functions for any bound state in a realistic Si-$SiO_2$ spherical quantum dot, including the material dependence of the electron effective mass. The influence of dot radius, confinement barrier potential and barrier-to-dot electron mass ratio on the electronic structure is investigated in detail. The results show that the energy structure shifts down from some tens to some hundreds meV compared to that obtained in the simplified model where the change in effective mass is neglected. Our exact single-electron calculation is finally used to verify the accuracy of the results obtained from a numerical approximate method developed to treat many-electron systems.

PACS numbers: 71.15.La, 73.20.Dx

Keywords: A. Semiconductors; A. Nanostructures; D. Electronic states

---

* Corresponding author. E-mail : philippe.dollfus@ief.u-psud.fr
Fax : +33 (0)1 69 15 40 20

## 1. Introduction

Efforts to integrate sub-10 nm-scaled CMOS devices will face physical limitations in the near future. To continue further increases in density with low power consumption, the trend is now to develop new device concepts turning the size reduction and quantum effects to advantage. In this field, one of the most promising idea is to use the Coulomb blockade mechanism in conducting island / tunnel junction systems to accurately control the current in single electron devices (SEDs). Self-assembled silicon nanocrystals (NCs) embedded in $SiO_2$ appear as a good basic structure for the realization of SEDs likely to operate at room temperature [1]. The compatibility of such NCs with Si technology makes them good candidates for actual applications in combination with conventional CMOS circuits. The concept of multi-dot memory using Si NCs as a floating gate has been demonstrated [2-4]. It could be developed to design nonvolatile multibit memory cells of very high scalability and reliability with low power consumption. The Coulomb blockade and quantization effects have been used in memory device with self-aligned doubly-stacked dots to improve the retention time [5] and demonstrated in single electron transistors (SETs) at room temperature [6].

The realisation and optimisation of SEDs requires accurate computational simulation of single electron charging effects to improve our understanding of the underlying physics. One important preliminary task is to calculate exactly the energy band structure inside the semiconductor NC where quantum effects become important. Several numerical techniques have been developed using either a tight-binding description [7] or an effective mass approximation [8-10]. The self-consistent Poisson-Schrödinger calculation based on the Hartree approximation or the density functional theory is convenient to determine the ground state of a many-electron system in the dot as in Ref. 10. Simple analytical models have been reported in the case of a single electron in a spherical quantum dot [11,12]. However, in these works the authors assumed the electron effective mass to be the same in the dot as in the surrounding barrier, which can make the results inappropriate to realistic applications. The present work deals with the exact calculation of single-electron energy and wave-function for all bound states in a realistic Si NC embedded in $SiO_2$ by including the change in effective mass from one material to the other. We give a detailed investigation of the energy structure for a wide range of dot radius, barrier height, and barrier-to-dot electron mass ratio. Our results are finally used to verify the accuracy of the results obtained using the numerical method developed in Ref. 10. It should be noted that the calculation is exact, fast and applicable for a spherical square-well quantum dot system of any size and potential barrier in any material system.

## 2. Model for exact calculation

We study the conduction band structure for a single-electron in a spherical QD within the model of finite square-well potential of radius $R$ and energy barrier $V_0$. Unless otherwise indicated, the effective atomic units are used in the whole article, i.e. the unit of mass is the free electron mass $m_0$, the unit of length is the effective Bohr radius, $a_B^{\,*} = a_B / m_{Si}$ (with $a_B$ = 0.529 Å) and the unit of energy is the effective Hartree $E_{Hart}^{\,*} = m_{Si}\, E_{Hart}$ (with $E_{Hart}$ = 27.212 eV). The electron effective mass in silicon $m_{Si}$ is approximated as the harmonic mean of the longitudinal and transverse masses in Δ valleys, i.e. $m_{Si} = 3\left(1/m_l + 2/m_t\right)^{-1} \approx 0.27$. On the basis of Weinberg experimental work [13] the barrier height $V_0$ and the electron effective mass in $SiO_2$ $m_{SiO_2}$ are chosen equal to 3.1 eV and 0.5, respectively. These values are frequently used in tunnelling calculations in the $Si/SiO_2$ system (see [14] and references therein).

In the effective mass approximation, which proves to be justified for the determination of the first levels even for dot radius as small as 15 Å [15], the single electron Schrödinger equation is given by

$$\left[-\frac{1}{2}\nabla_{\vec{r}}\left(\frac{1}{m^*}\nabla_{\vec{r}}\right)+V(\vec{r})\right]\Psi(\vec{r})=E\,\Psi(\vec{r}) \tag{1}$$

where

$$V(\vec{r})=\begin{cases}0 & ,\, r<R\\ V_0 & ,\, r\geq R\end{cases} \qquad \text{and} \qquad m^*=\begin{cases}m_{Si} & ,\, r<R\\ m_{SiO_2} & ,\, r\geq R\end{cases} \tag{2}$$

Due to spherical symmetry of the system, the wavefunction may be written in the form

$$\Psi(\vec{r})=Y_{l,m}(\varphi,\theta)\,F(r) \tag{3}$$

where $Y_{l,m}(\varphi,\theta)$ is a spherical harmonic with orbital quantum number $l$ ($l$ = 0,1,2,….) and azimuthal quantum number $m$ ($-l \leq m \leq l$).

It can be easily shown from Eq. (1) and (2) that the differential equations for the radial part $F(r)$ inside and outside the dot are just the equations of spherical ordinary and modified Bessel functions, $j_l(r)$ and $k_l(r)$, respectively. The solutions, which are finite at the center of the dot ($r$ = 0) and tend to zero when $r$ increases indefinitely are given by

- inside the dot:

$$F(r) = A\, j_l(\alpha r) = A\sqrt{\frac{\pi}{2\alpha r}}\, J_{l+\frac{1}{2}}(\alpha r)\ ; \quad \text{with} \quad \alpha = \sqrt{E} \tag{4}$$

- outside the dot:

$$F(r) = B\, k_l(\beta r) = B\sqrt{\frac{\pi}{2\beta r}}\, K_{l+\frac{1}{2}}(\beta r)\ ; \quad \text{with } \beta = \sqrt{\gamma(V_0 - E)},\ \gamma = \frac{m_{SiO_2}}{m_{Si}} \tag{5}$$

where $J_{l+\frac{1}{2}}(\alpha r)$ is the Bessel function of the first kind and $K_{l+\frac{1}{2}}(\beta r)$ is the modified Bessel function of the second kind. The normalization coefficient $B$ is defined from the continuity condition imposed on the wavefunction at the dot surface: $B = A\, j_l(\alpha R)/k_l(\beta R)$. From the normalization of the wavefunction, $A$ is given by

$$A = (T_1 + T_2)^{-1/2}\text{, where } T_1 = \int_0^R [j_l(\alpha r)]^2\, r^2\, dr,\ T_2 = \left|\frac{j_l(\alpha R)}{k_l(\beta R)}\right|^2 \int_R^\infty [k_l(\beta r)]^2\, r^2\, dr \tag{6}$$

To determine the energy levels in the dot, we use the transcendental equation resulting from the boundary condition imposed on the derivative of the wave-function at the dot surface, i.e., by taking into account the change in effective mass between Si and $SiO_2$,

$$j_l'(\alpha R) = \frac{1}{\gamma}\frac{j_l(\alpha R)}{k_l(\beta R)}\, k_l'(\beta R) \tag{7}$$

Using the following formulas for derivatives

$$j_l'(r) = j_{l-1}(r) - \frac{l+1}{r} j_l(r)\ ; \quad \text{and } k_l' = -k_{l-1}(r) - \frac{l+1}{r} k_l(r) \tag{8}$$

we get the following equation from which the discrete values of electron energy $E = \alpha^2 = V_0 - \beta^2/\gamma$ can be obtained:

$$\gamma\left[\frac{\alpha\, j_{l-1}(\alpha R)}{j_l(\alpha R)} - \frac{(l+1)}{R}\right] = -\frac{\beta\, k_{l-1}\left(\sqrt{\gamma(V_0 - \alpha^2)}\, R\right)}{k_l\left(\sqrt{\gamma(V_0 - \alpha^2)}\, R\right)} - \frac{(l+1)}{R} \tag{9}$$

Note that, if we ignore the difference of electron effective mass between the different materials, the equation (9) simplifies to

$$\frac{\alpha\, j_{l-1}(\alpha R)}{j_l(\alpha R)} = -\frac{\beta\, k_{l-1}(\beta R)}{k_l(\beta R)}\ , \ \text{with } \alpha^2 + \beta^2 = V_0 \tag{10}$$

which has been used by various authors [11-12].

Once the energies $E_{l,n} = \alpha_{l,n}^2$ which satisfy Eq. (9) are determined, the exact wave-functions of bound states are given by

$$\Psi_{l,m,n}(\vec{r})=\begin{cases} A\, j_l(\alpha_{l,n}\, r)\, Y_{l,m}(\varphi,\theta) & , r < R \\ A\, \dfrac{j_l(\alpha_{l,n}\, R)}{k_l\left(\sqrt{\gamma[V_0-\alpha_{l,n}{}^2]}\; R\right)}\, k_l\left(\sqrt{\gamma[V_0-\alpha_{l,n}{}^2]}\; r\right) Y_{l,m}(\varphi,\theta) & , r \geq R \end{cases} \tag{11}$$

with $A$ defined by Eq. (6)

Given the radius $R$, the confinement barrier height $V_0$ and the mass ratio $\gamma$, Eq. (9) is solved to determine all allowed bound states $\alpha_{l,n}$ in the dot. The solution has been carried out by computation code written in *Mathematica* environment [16], yielding high accuracy within manageable calculation time.

## 4. Results and discussion

We are interested in spherical Si-$SiO_2$ quantum dots with typical radius $R$ in the range 10 Å-60 Å. The barrier energy $V_0$ is 3.1 eV, and the mass ratio is $\gamma = 1.85$. However, for purposes of comparison and to analyse the influence of these parameters, $V_0$ and $\gamma$ have been considered to vary in the range 1 eV-30 eV and 0.1-2.5, respectively.

We plot in Fig. 1 the $R$ dependence of the 1s-level for $\gamma = 1$ and for $V_0 = 1$ eV, 3.1 eV and 30 eV. In Fig. 2 the $V_0$ dependence of the same level is shown for $\gamma = 1$ and $R = 8$ Å, 1 Å, 15 Å, 30 Å. These figures verify that the energy increases when the radius decreases and when the barrier potential increases, which is a common feature of quantum confinement. Note that the dependence on this parameters is all the stronger that the radius and the barrier energy become smaller.

To see the effect of the mass difference between the dot and the surrounding barrier we depict in Fig. 3 the radius dependence of 1s-level for $V_0 = 3.1$ eV $\gamma = 0.5$, 1, 2.5. It shows that the energy decreases when the mass ratio is larger. A similar feature is observed in Fig. 4 for the barrier potential dependence of the 1s-level ($R$ = 30Å, and $\gamma = 0.5$, 1, 1.85). We still see that at given $R$ and $V_0$ the energy is lower when $\gamma$ increases; moreover, the change of energy as a function of $\gamma$ is all the higher that $V_0$ is smaller.

To estimate quantitatively the change of the electron energy depending on whether the difference of effective mass between materials is taken into account or not, we plot in Figs. 5 and 6 the quantity $E_{\gamma=1} - E_{\gamma=1.85}$ as a function of dot radius and barrier potential, respectively. The three first levels 1s, 2p and 3d are considered in these graphs. We can see clearly that the magnitude of the energy change is larger for small dots and small barrier energies. The typical energy change is about some tens meV for large dots and reaches more than 100 meV for small dots, which cannot be neglected for most practical applications.

Despite the effective mass approximation should fail at high energy in Si, we plot in Fig. 7 the full picture of the energy structure of a single electron quantum dot Si-$SiO_2$ as a function of the principal quantum number $n$ (with $\gamma = 1$, $V_0 = 3.1$ eV and $R = 30$ Å). It is also plotted in Fig. 8 as a function of the orbital number $l$ (with $V_0 = 3.1$ eV and $R = 30$ Å, and both $\gamma = 1$ and $\gamma = 1.85$). If we use the conventional spectroscopic notation for the electron energy levels, we obtain an order of energy levels quite different from that of the hydrogen atom. The sequence is as follows

$$E_{1s} < E_{2p} < E_{3d} < E_{2s} < E_{4f} < E_{3p} < E_{5g} < E_{4d} < E_{6h} < E_{3s} < ...$$

The difference in the order of levels between the hydrogen atom and the semiconductor QD is explained by the fact that in hydrogen atom the Coulomb potential is deeply singular at $r = 0$, while there is no singularity in the dot. This sequence agrees with the results reported in Ref. 11. We can also note that the separation of the successive energy levels increases when the number $n$ increases, i.e. when we go up to the top of the dot well. This feature is quite contrary to the case of a normal atom, in which the separation between energy levels of highly excited states becomes smaller and smaller when $n$ increases.

In Fig. 8 we can compare the energy structure in two cases: by including the material-dependence of mass in the calculation (dashed lines) or not (solid line). In accordance with results presented above, the energy levels resulting from a realistic model ($\gamma = 1.85$) are lower than using the simplified model ($\gamma = 1$). This situation may result in the emergence of new bound states which have been omitted in the simplified calculation. For example, the highly excited level 6d and 9k in Fig. 8 are only observed for $\gamma = 1.85$. Additionally, the energy shift being significantly dependent on $l$ value, one may observe a change in the sequence of levels (see e.g. 11*n* and 7*g*). By choosing the appropriate radius, barrier energy and mass ratio, one can expect this situation to happen for lower levels.

Finally, we would like to compare our exact calculation with the results obtained from an approximate numerical method developed to treat many electron systems. Fig. 9 presents the radius dependence of the first energy level calculated either by our exact treatment (solid line) or by self-consistently solving Poisson-Schrödinger equations using the Hartree approximation [10] (dashed line). In Fig. 10 the corresponding wave functions resulting from both methods are shown. The excellent agreement obtained for the ground state is observed for higher levels as well (not shown), confirming the validity of both methods.

## Conclusion

We have presented the exact calculation of energy structure of a single electron in Si spherical nanocrystal surrounded by $SiO_2$ forming a square well with finite barrier potential. Contrary to the case of simplified models used so far, the present model takes into account the material dependence of the electron effective mass. Fast and accurate calculations can be carried out to generate the energy structure for any combination of three inputs: radius of the dot, barrier potential of the dot and the ratio of barrier to dot electron effective mass. It is shown that the energy resulting from the simplified approach (with $\gamma = 1$) is overestimated for all states. The effect of material-dependence of electron effective mass becomes all the more significant that the quantum dot size and the barrier energy are smaller: the difference between realistic and simplified calculations of energy levels may reach more than 100 meV in small dots. Additionally, our calculation may result in new energy levels omitted by the simplified method and in a modified order of energy levels. As a future extension of this work, the basis of Bessel-like wavefunctions determined for a single electron could be generalised to treat the effect of Coulomb interaction in many-electron quantum dots. Such an approach should be faster to compute than the usual self-consistent Poisson-Schrödinger solver.

## Acknowledgement

One of the authors (N. H. Quang) wishes to acknowledge financial support of the CNRS (PICS program). He is most appreciative for the hospitality shown to him while a visitor at the IEF, Orsay.

## Figure Captions

Figure 1. Dependence of the 1s-energy level on the dot radius $R$ for some values of barrier height $V_0$ ($\gamma = 1$).

Figure 2. Dependence of the 1s-energy level on the confinement potential $V_0$ for some values of radius $R$ ($\gamma = 1$). The vertical dashed line shows the value $V_0$ = 3.1 eV.

Figure 3. Dependence of the 1s-energy level on the dot radius $R$ for some values of mass ratio $\gamma$ ($V_0$ = 3.1 eV).

Figure 4. Dependence of the 1s-energy level on the confinement potential $V_0$ for some values of mass ratio $\gamma$ ($R$ = 30 Å).

Figure 5. Change in 1s, 2p and 3d energy levels when the material-dependence of effective mass is included in the calculation. Results are plotted as a function of dot radius $R$ at given $V_0$ = 3.1 eV.

Figure 6. Change in 1s, 2p and 3d energy levels when the material-dependence of effective mass is included in the calculation. Results are plotted as a function of barrier height $V_0$ at given $R$ = 30 Å.

Figure 7. Order of all energy levels in the QD for $V_0$ = 3.1 eV, $R$ = 30 Å and $\gamma = 1$.

Figure 8. Order of all energy levels in the QD for $\gamma = 1$ (solid lines) and $\gamma = 1.85$ (dashed lines) ($V_0$ = 3.1 eV, $R$ = 30 Å).

Figure 9. 1s energy level as a function of dot radius. Comparison of exact calculation with Poisson-Schrödinger solution of Ref. 10 ($\gamma = 1.85$, $V_0$ = 3.1 eV).

Figure 10. Squared wave function of 1s level. Comparison of exact calculation with Poisson-Schrödinger solution of Ref. 10 ($\gamma = 1.85$, $V_0$ = 3.1 eV, R = 30 Å).

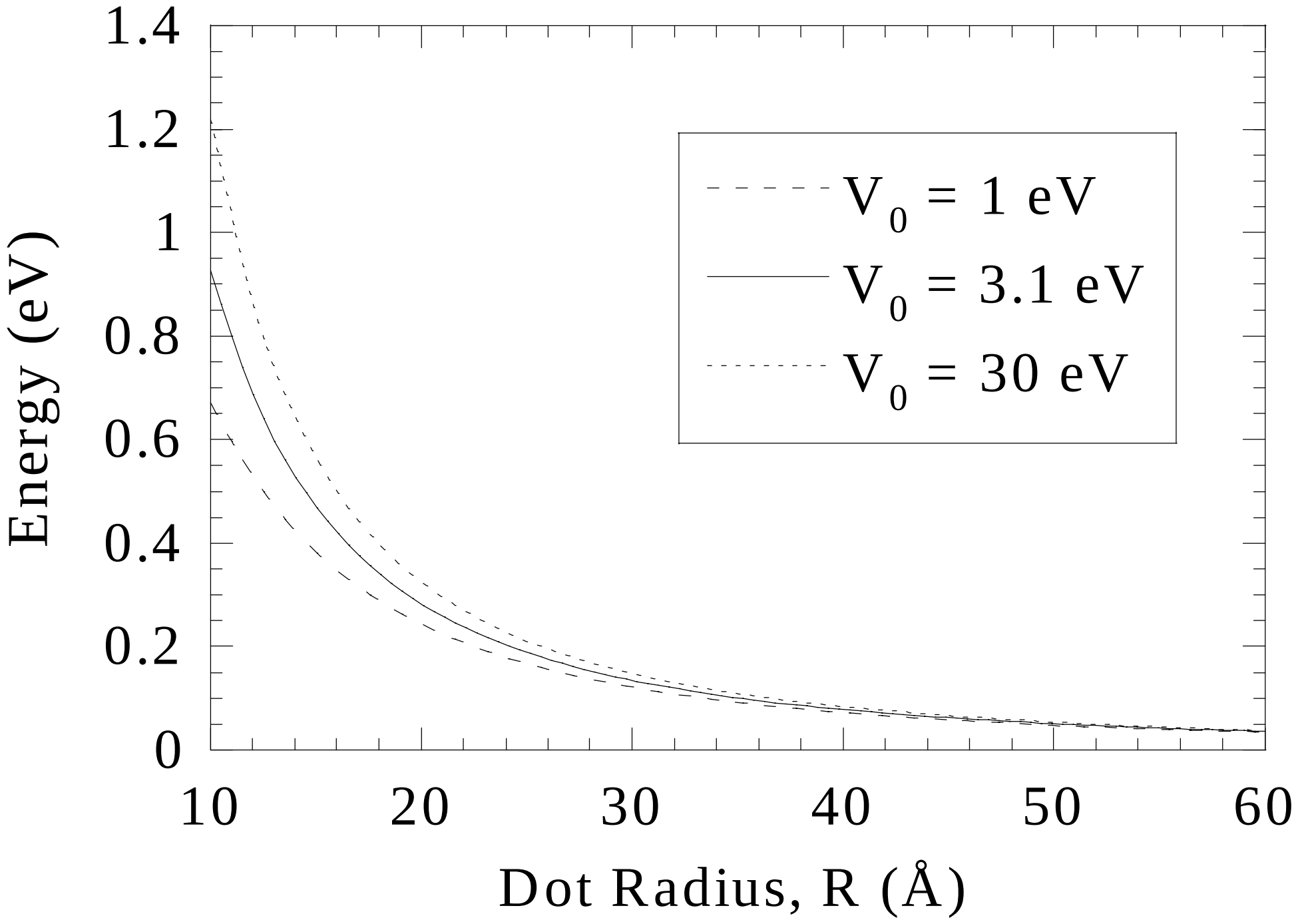
1.4
1.2
1
0.8
0.6
0.4
0.2
0
Energy (eV)
V0 = 1 eV
V0 = 3.1 eV
V0 = 30 eV
10
20
30
40
50
60
Dot Radius, R (Å)

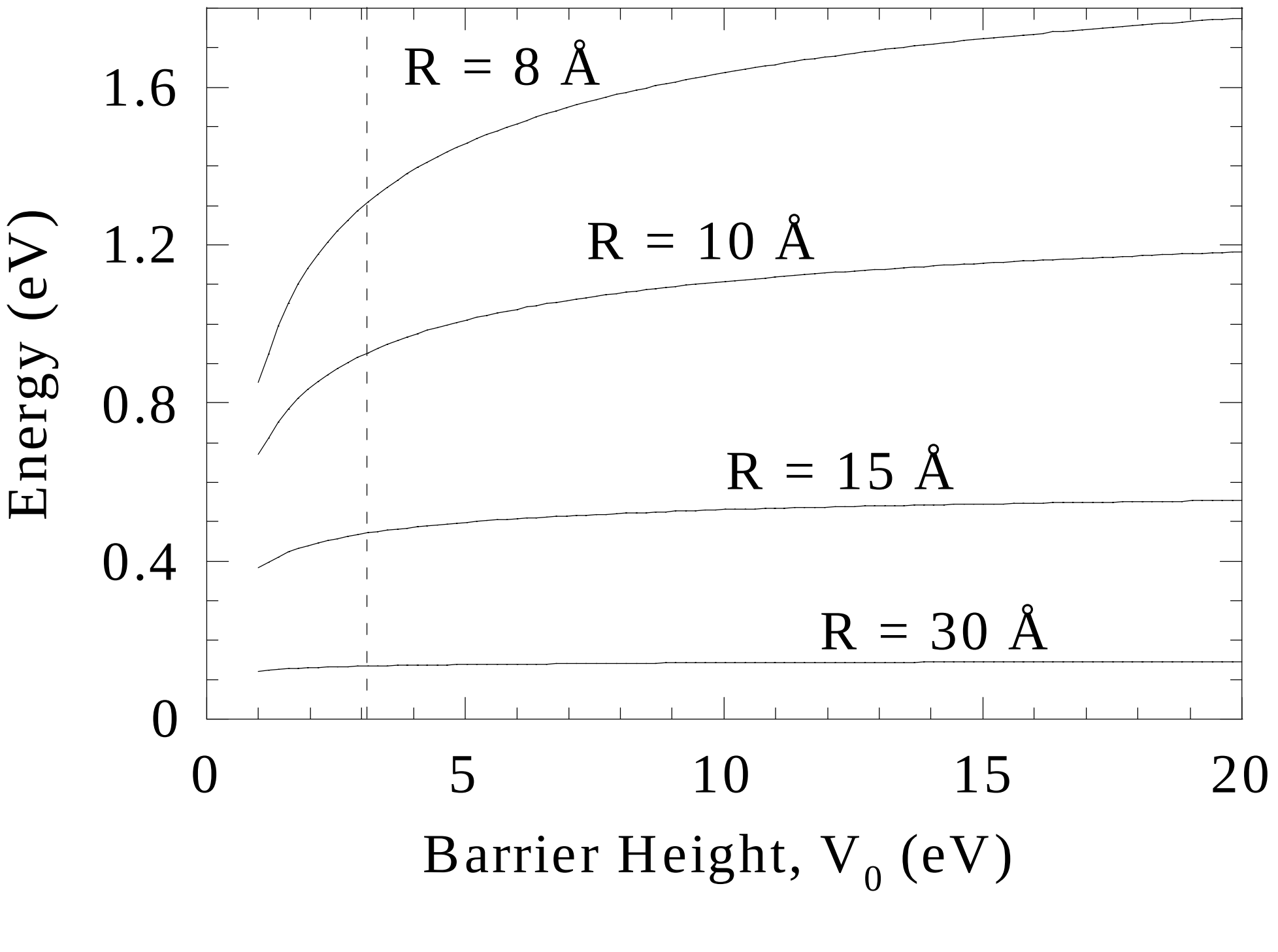
R = 8 Å
R = 10 Å
R = 15 Å
R = 30 Å
Energy (eV)
1.6
1.2
0.8
0.4
0
0
5
10
15
20
Barrier Height, V0 (eV)

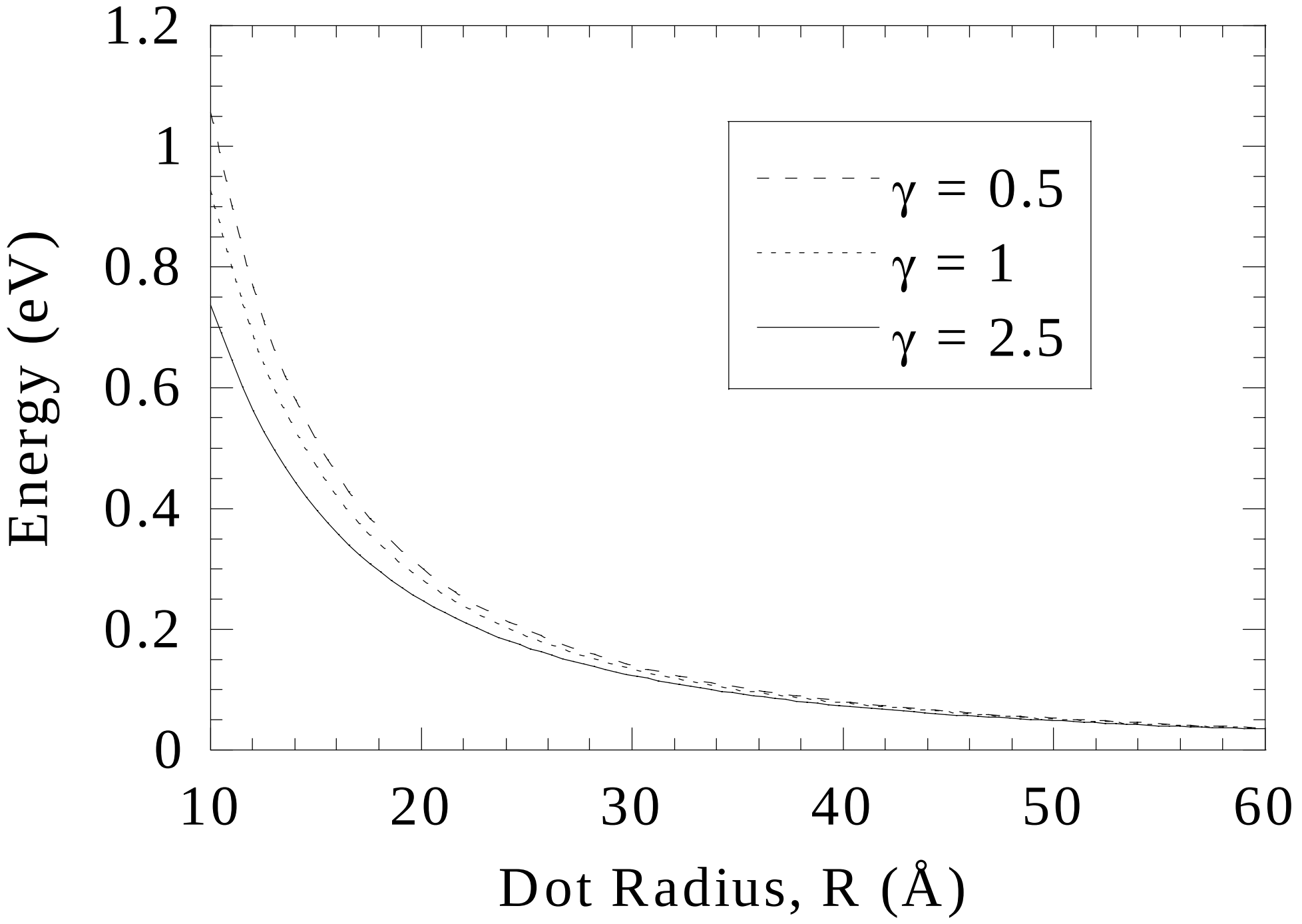

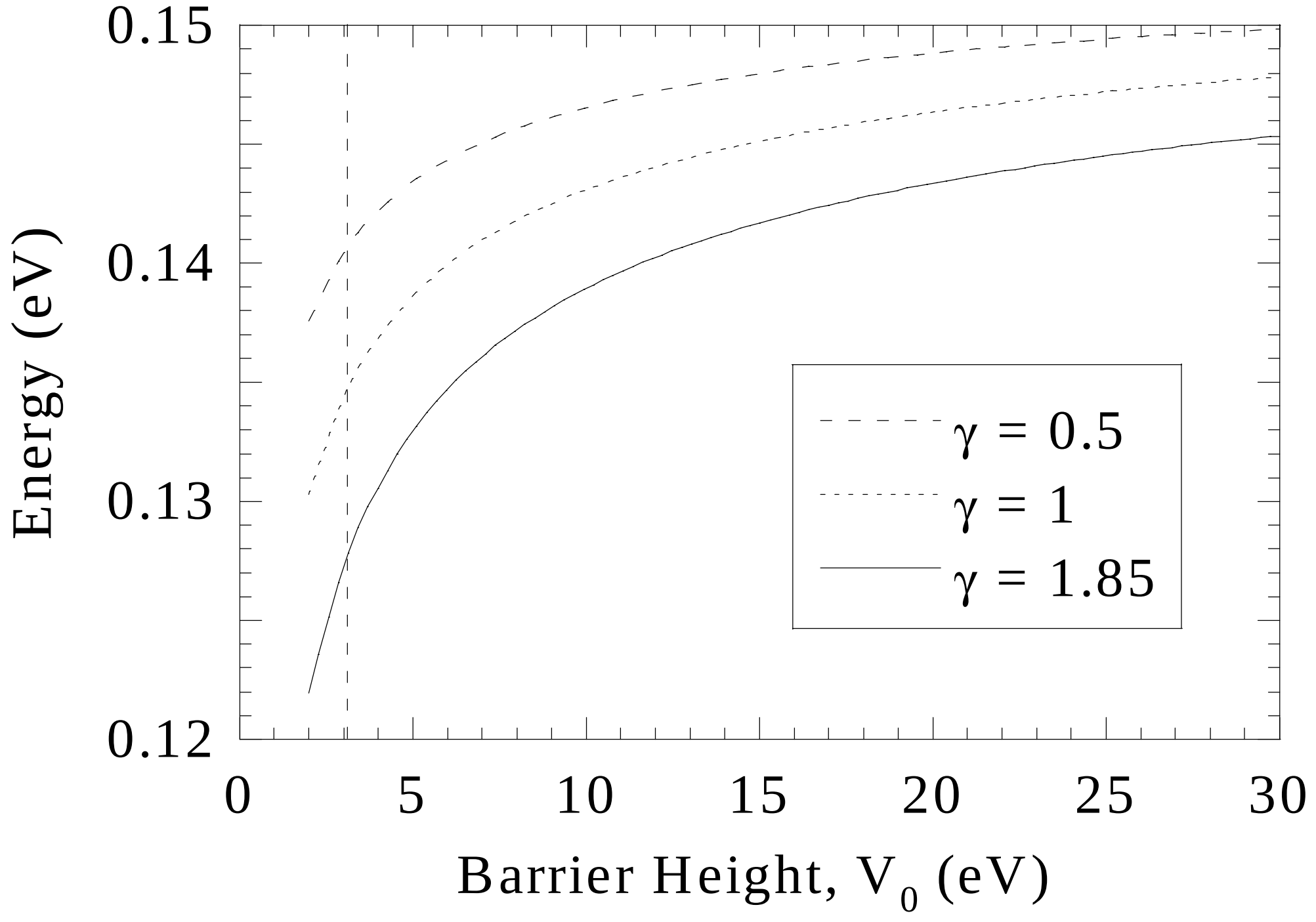

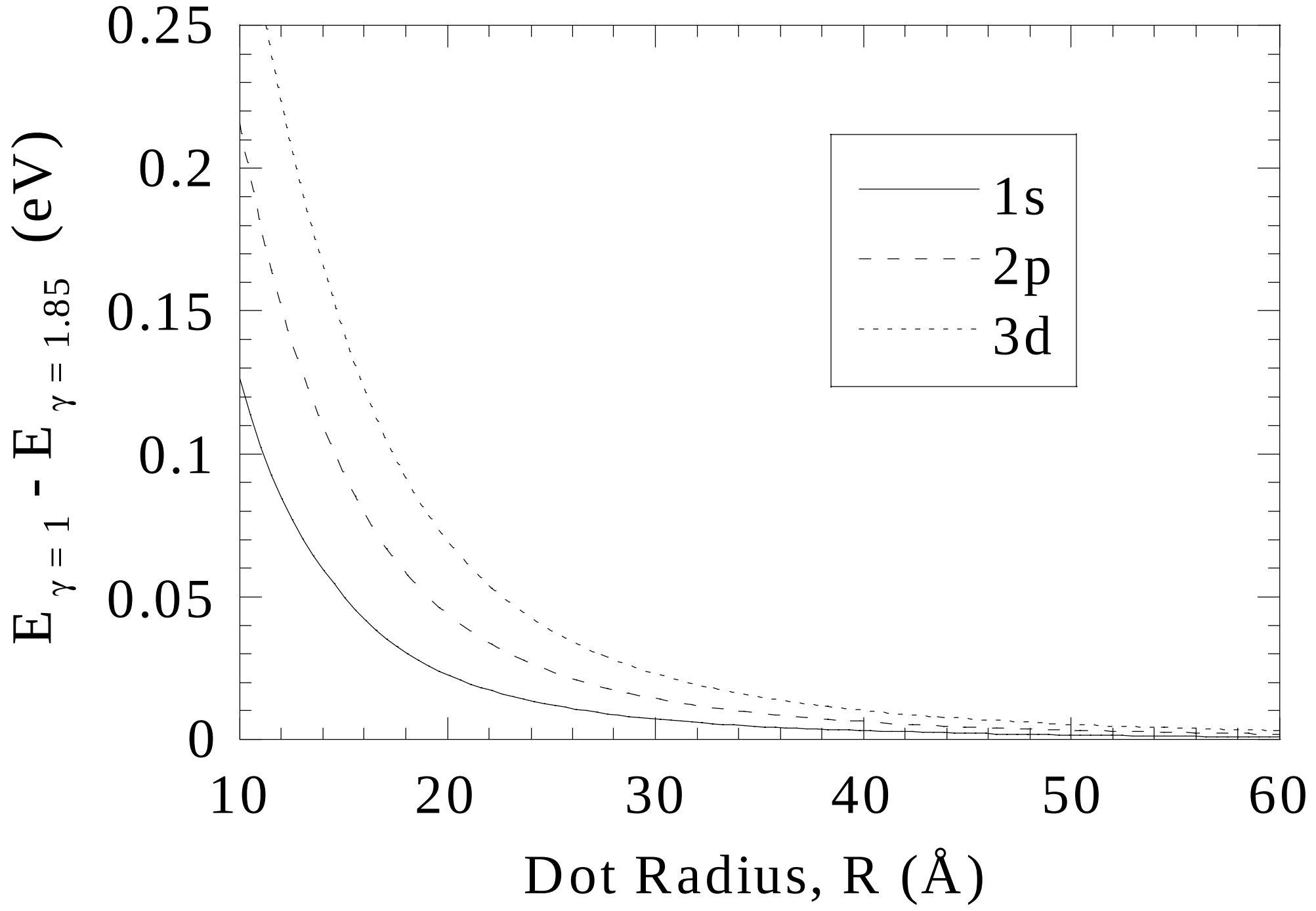
0.25
0.2
0.15
0.1
0.05
0
10
20
30
40
50
60
Dot Radius, R (Å)
E γ = 1 - E γ = 1.85 (eV)
1s
2p
3d

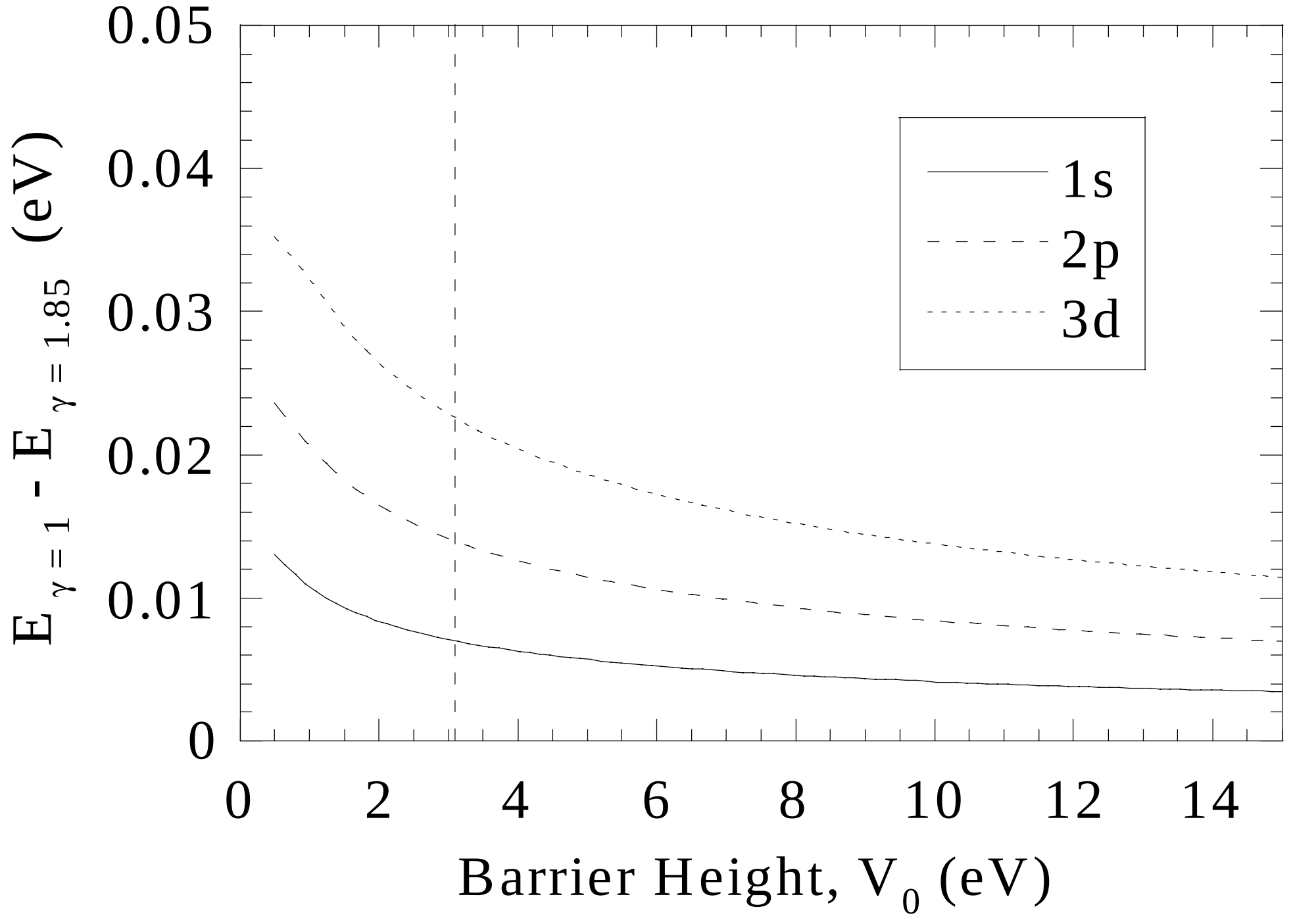
0.05
0.04
0.03
0.02
0.01
0
0
2
4
6
8
10
12
14
1s
2p
3d
E γ = 1 - E γ = 1.85 (eV)
Barrier Height, V0 (eV)

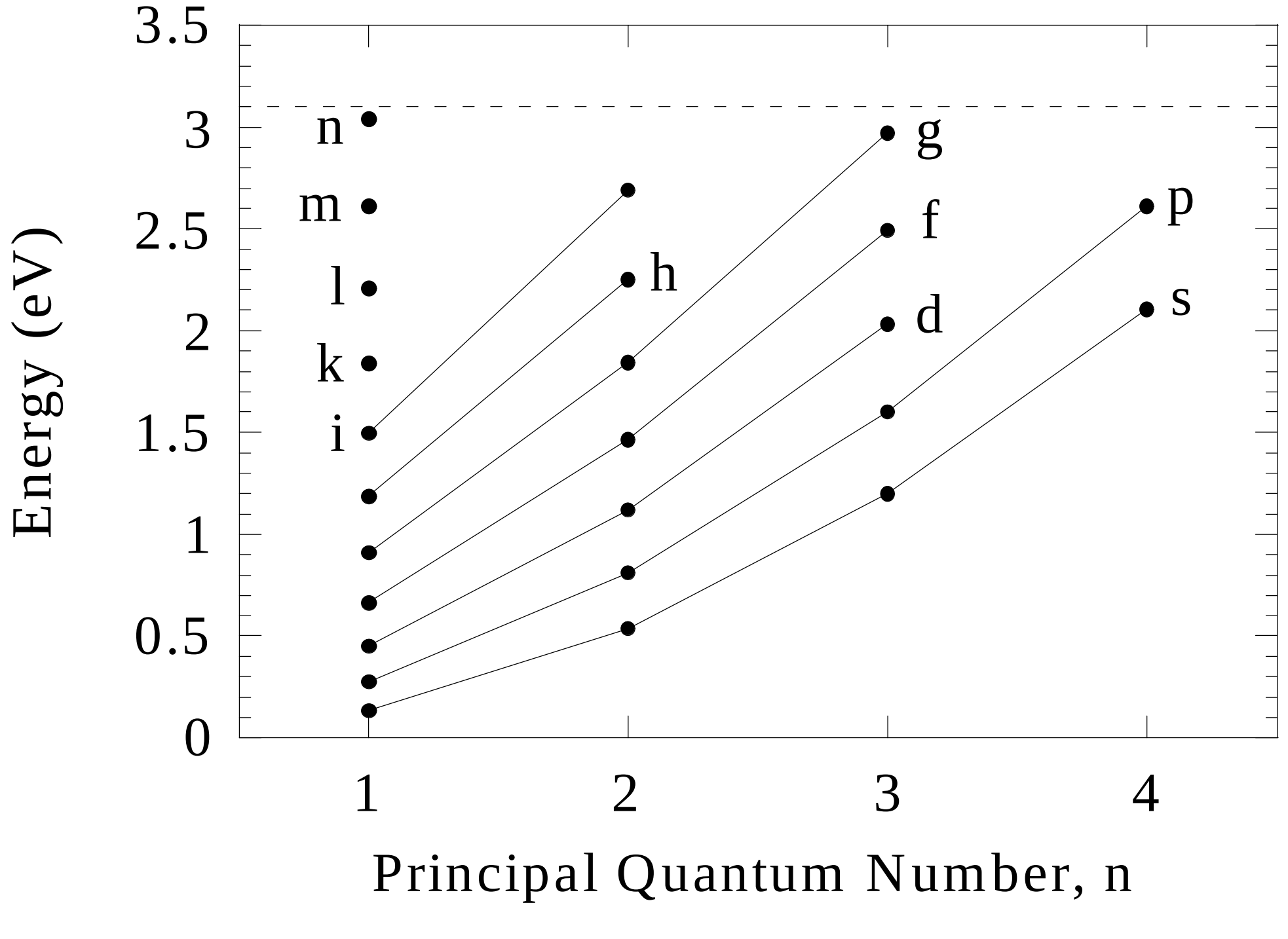

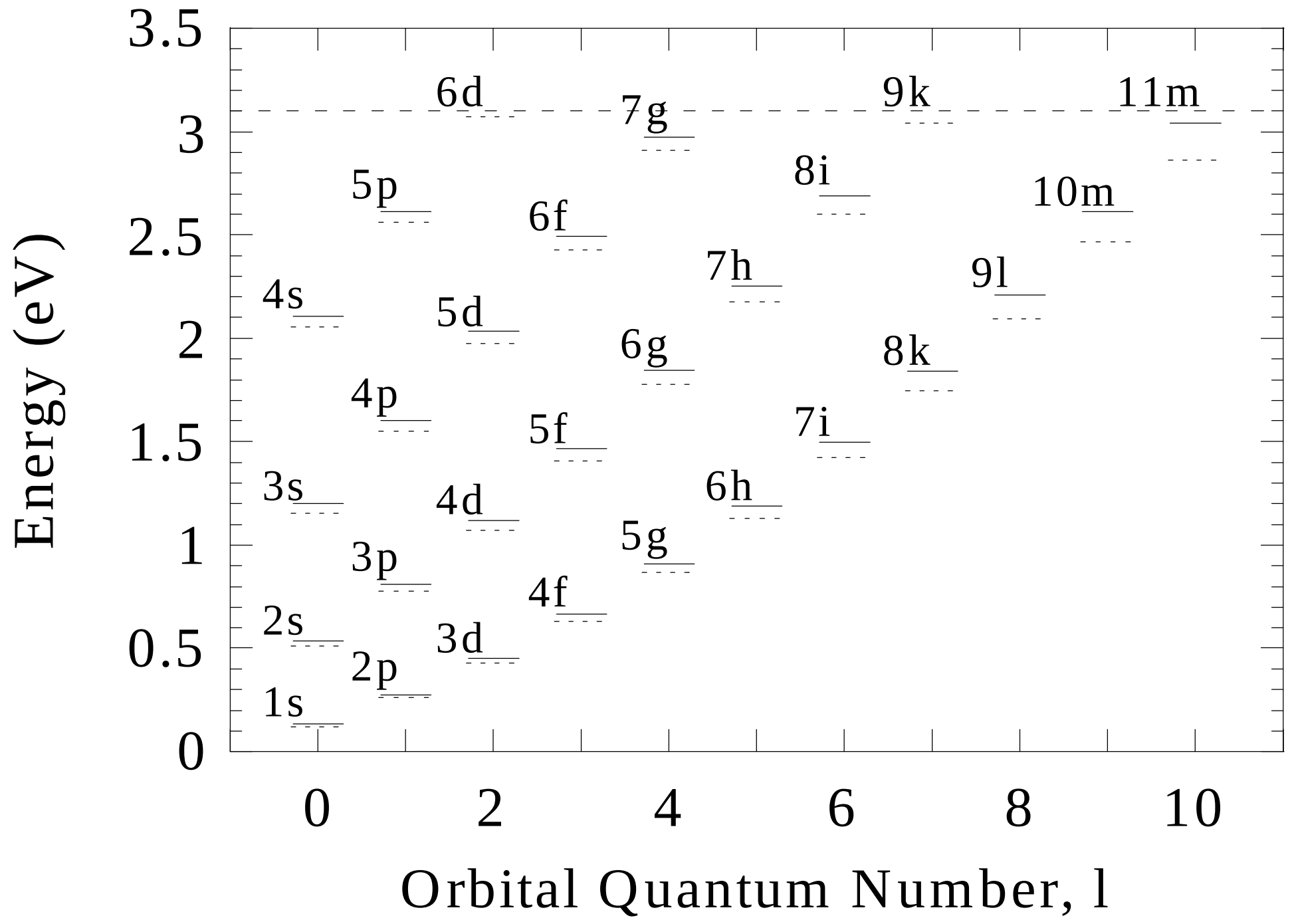

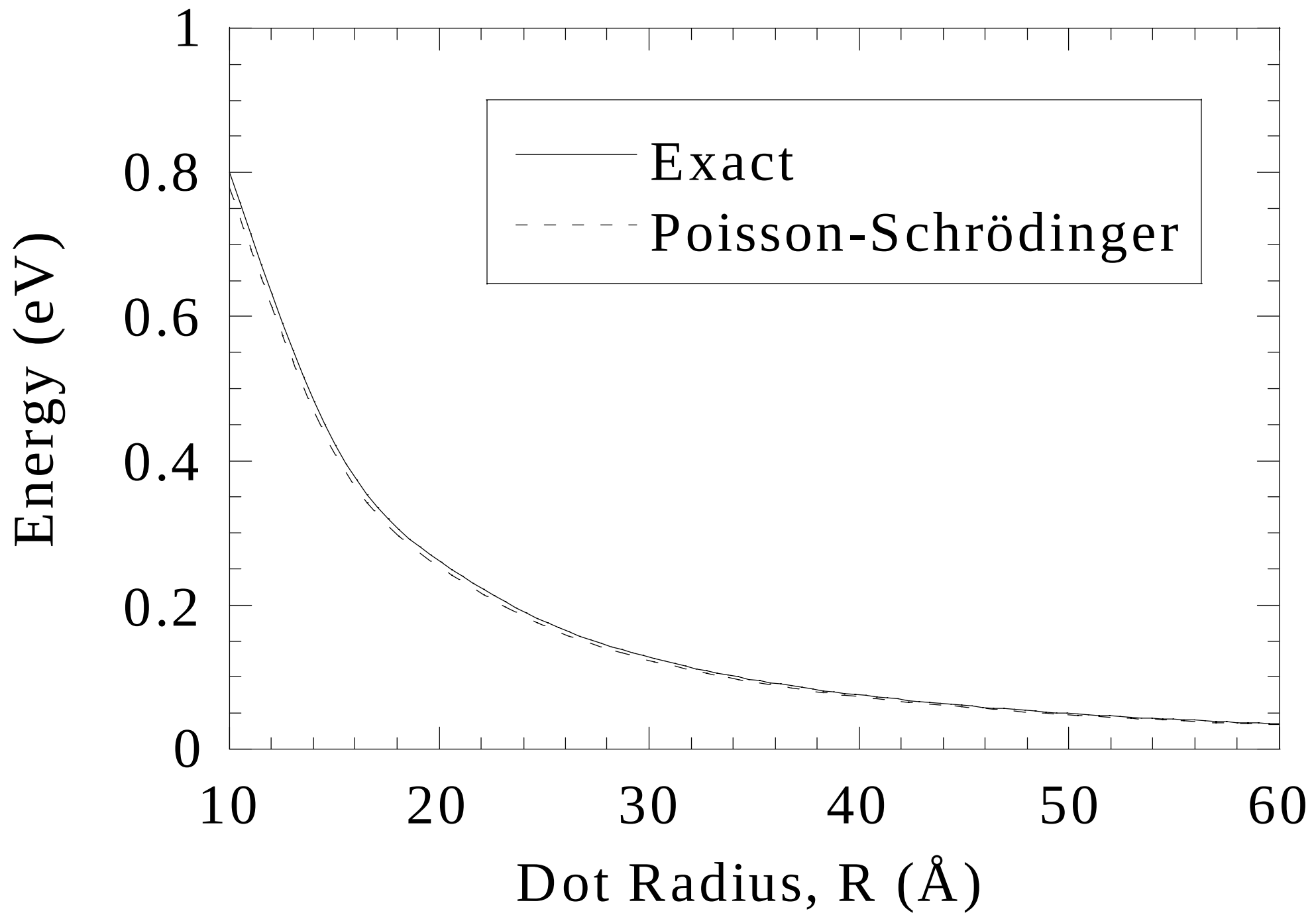

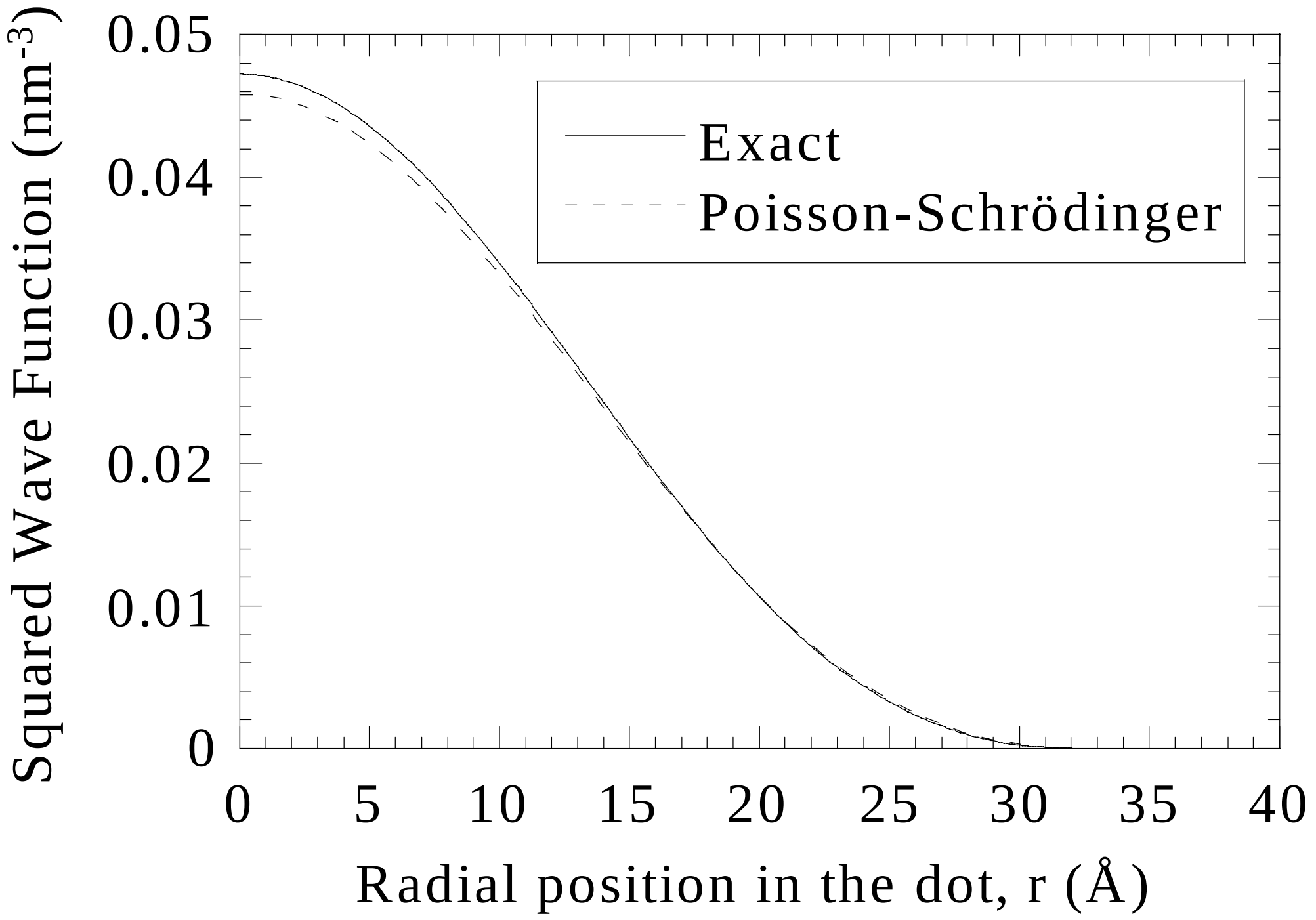